\documentclass{optica-article}

\journal{opticajournal} 
\articletype{Research Article}

\begin{document}

\title{Magneto-optical intensity effect in a transparent perforated all-dielectric metasurface with an anapole state}

\author{Polina V. Zorina\authormark{1,2,3,*}, Andrey N. Kalish\authormark{3}, Vladimir I. Belotelov\authormark{2,3}, Venu Gopal Achanta\authormark{4,5}, and Daria O. Ignatyeva\authormark{2,3}}

\address{\authormark{1}Moscow Institute of Physics and Technology (National Research University) (MIPT), Moscow 141701, Russia\\
\authormark{2}Faculty of Physics, Lomonosov Moscow State University, Moscow 119992, Russia\\
\authormark{3}Russian Quantum Center (RQC) (International Center for Quantum Optics and Quantum Technologies), Moscow 141026, Russia\\
\authormark{4}Department of Condensed Matter Physics and Materials Science, Tata Institute of Fundamental Research, Mumbai 400005, India\\
\authormark{5}National Physical Laboratory, Dr. K. S. Krishnan Marg, New Delhi 110012, India}

\email{\authormark{*}p.zorina@rqc.ru}

\begin{abstract*}
We numerically study a perforated all-dielectric metasurface based on the magnetic semiconductor GaMnAs. Near the anapole state, the structure combines high transmittance with a moderate magneto-optical intensity effect. The absolute transmission modulation is enhanced by about a factor of five compared with a smooth film, while the metasurface transmittance reaches 81\%. The enhancement is associated with strong field localization in the magneto-optically active material and suppression of the electric-dipole radiation channel in the anapole regime.
\end{abstract*}



\section{Introduction}

Magneto-optical effects provide a powerful route to controlling light intensity, phase, and polarization through an external magnetic field or the magnetization of a medium. They are widely used in optical modulators, nonreciprocal photonic elements, and polarization-control devices. In thin films and nanostructures, however, the magnitude of magneto-optical effects is often limited by the short light--matter interaction length. A key challenge is therefore to enhance the magneto-optical response while preserving a sufficiently high intensity of the transmitted optical signal.

One route toward enhancing the magneto-optical response is to nanostructure the medium so that the optical response is governed not only by the material parameters but also by the resonant modes and field distribution of the structure. This principle is used, for example, in integrated photonic structures for analog optical-signal processing \cite{Kashapov2024Differentiation} and in three-dimensional microstructures for efficient coupling of light into photonic integrated circuits \cite{Kolymagin2024Microstructures}. Nanoscale magnetophotonics combines resonantly localized optical fields with magnetic functionality, enabling control over light intensity, polarization, and phase \cite{Maccaferri2020Nanoscale}.

Several resonant mechanisms have been employed to strengthen the interaction of light with magnetic media. In periodic magnetic structures, slow-light conditions near photonic-band edges can enhance the Faraday effect by increasing the effective light--matter interaction \cite{Belotelov2009SlowLight}. In one-dimensional magneto-optical photonic crystals, defect-mode localization can strongly enhance Faraday rotation, while the spectral position of its maximum may depend on the wavelength-dependent off-diagonal dielectric response and optical absorption \cite{Inui2011Faraday}. Hybrid magnetophotonic--plasmonic structures can additionally combine microresonator, waveguide, and plasmonic modes to enhance Faraday and Kerr effects \cite{Khokhlov2015PhotonicCrystals}. In ferromagnetic nanoantennas, the phase of localized plasmon resonances can be used to tune the spectral shape and sign of the magneto-optical response \cite{Maccaferri2013NiDisks}. Related concepts have also been implemented in hybrid structures combining dielectric Mie resonators with magnetic layers, magnetoplasmonic systems, and all-dielectric metasurfaces \cite{Barsukova2017Mie,Kharratian2021Faraday,Voronov2020Subwavelength}. High-index dielectric resonators are especially attractive because they support electric and magnetic multipolar modes and provide strong light--matter interaction with lower optical losses than plasmonic structures. As a result, nanostructuring a magneto-optical material, or coupling it to dielectric resonators, can lead to a substantial enhancement of the magneto-optical response and even give rise to effects that are absent or negligible in planar films \cite{Xia2022Circular}. Magneto-optical iron garnet films are also actively studied, including epitaxial films and films patterned with surface arrays of ferromagnetic metal particles. Such systems have been used for magnetic-domain imaging by second- and third-harmonic generation, as well as for studying how surface metastructures affect the nonlinear optical response and domain-wall pinning \cite{Kolmychek2025IronGarnet}. This highlights the broader role of artificial structuring in magnetic photonic systems, where geometry can affect both the optical response and the magnetic state of the material.

A magnetized medium in the Faraday geometry exhibits different optical responses to right- and left-circularly polarized light. In the conventional absorption-based description, this response is known as magnetic circular dichroism and is associated with a difference in the absorption coefficients for the two circular polarizations. In metasurfaces operating in transmission, the same gyrotropic response can manifest itself as a difference in transmittance. This effect should be distinguished from circular dichroism in chiral metasurfaces, where the asymmetry between right- and left-circularly polarized light is induced by the geometry itself, for example by chiral meta-atoms, composite unit cells, or broken mirror symmetries of the lattice \cite{Wang2023CircularDichroism,Ali2023Chiral,Toftul2024Chiral}. In a magneto-optical resonator, the polarization-dependent response is governed by the magnetization and the gyration parameter of the material, but it is also shaped by the resonant modes supported by the structure. This is consistent with multipolar analyses of magneto-dielectric scatterers, where magnetic circular dichroism is shown to depend on individual multipole contributions \cite{HernandezSarria2026MCD}. At the same time, resonant enhancement of magneto-optical effects is often accompanied by increased scattering, reflection, or absorption, which reduces the transmitted signal. Therefore, for the magneto-optical intensity effect in transmission it is important to employ resonant states that combine field localization inside the structure with high transmittance.

All-dielectric nanophotonics provides a suitable platform for realizing resonant states that combine strong field localization with weak radiative scattering. In contrast to plasmonic systems, high-index dielectric nanostructures support electric and magnetic multipolar resonances associated with displacement-current distributions inside the dielectric volume. Their multipolar description can also be extended to toroidal contributions and their higher-order counterparts, associated with vortex-like displacement-current configurations, providing additional degrees of freedom for controlling scattering, near fields, and light--matter interaction at comparatively low optical losses \cite{Gurvitz2019Toroidal}.

A prominent example of a weakly radiating resonant state is the optical anapole. Although the concept was originally introduced by Ya. B. Zel'dovich in elementary-particle physics, in nanophotonics it describes a configuration in which far-field radiation is strongly suppressed. In its simplest form, this suppression results from destructive interference between electric-dipole and toroidal-dipole contributions, which have the same far-field radiation pattern and can cancel for an appropriate relation between their amplitudes and phases. Consequently, the dipolar far-field response is reduced while the electromagnetic field remains strongly localized inside the structure; in anapole metamaterials, this mechanism can produce resonances with extremely high quality factors \cite{Basharin2017Anapole}.

This combination of suppressed scattering and enhanced near fields makes anapole states attractive for enhancing optical effects. Beyond the conventional electric-dipole anapole, high-index dielectric particles can also support magnetic anapoles and hybrid regimes in which the electric- and magnetic-dipole radiation channels are simultaneously suppressed \cite{Lukyanchuk2017HybridAnapole}. Anapole-type near-field interference can also be used to engineer the spatial distribution of the local electric and magnetic fields; coupled high-index dielectric tubes have been shown to provide pronounced magnetoelectric field separation \cite{Baryshnikova2018FieldSeparation}. In dielectric nanostructures, anapole-related scattering minima can be accompanied by substantial electric-field enhancement. This has been demonstrated in individual silicon nanodisks and modified slotted geometries, where anapole excitation leads to pronounced field concentration while maintaining a weakly scattering response \cite{Yang2018Anapole}.

Anapole states have most commonly been studied in individual dielectric disks or arrays of such resonators. For nonlinear and magneto-optical effects, however, this geometry is not always optimal, because replacing a continuous film with discrete resonators reduces the amount of active material participating in light--matter interaction. A perforated film provides an alternative design: most of the functional material is preserved, while the holes shape the resonant field distribution required for anapole excitation. This geometry is therefore especially relevant for enhancing the magneto-optical response in structures where the structured material itself is magneto-optically active.

High-index metasurfaces with arrays of circular nanoholes have been shown to support anapole states in which suppressed scattering coexists with high transmittance, while the associated field localization can strongly enhance the effective Kerr nonlinearity \cite{Panov2022Kerr}. Perforated magnetic films are also of interest beyond their optical response, since the hole geometry can affect the magnetic configuration; in particular, electric-field control of the magnetic-structure topology has been demonstrated for a planar ferromagnetic film \cite{Magadeev2025Perforated}.

In our previous study of a hybrid silicon--garnet metasurface \cite{Zorina2026Helicity}, the resonant field associated with the anapole state was localized mainly inside the nonmagnetic silicon nanodisks and only weakly overlapped with the underlying magnetic layer, limiting the magneto-optical enhancement. In contrast to such hybrid dielectric--magnetic metasurfaces, where the resonant field can be predominantly confined within a nonmagnetic resonator, here the resonant state is supported directly by the magneto-optically active material. The perforated GaMnAs film thus simultaneously supports the anapole state and provides the magneto-optical response. We show that this architecture enables resonant field localization directly within the magnetic material while preserving high transmittance and enhancing the magneto-optical intensity effect.

\section{Model and computational methods}

\subsection{Metasurface geometry and material model}

\begin{figure}[htbp]
\centering
a)\includegraphics[width=0.55\linewidth]{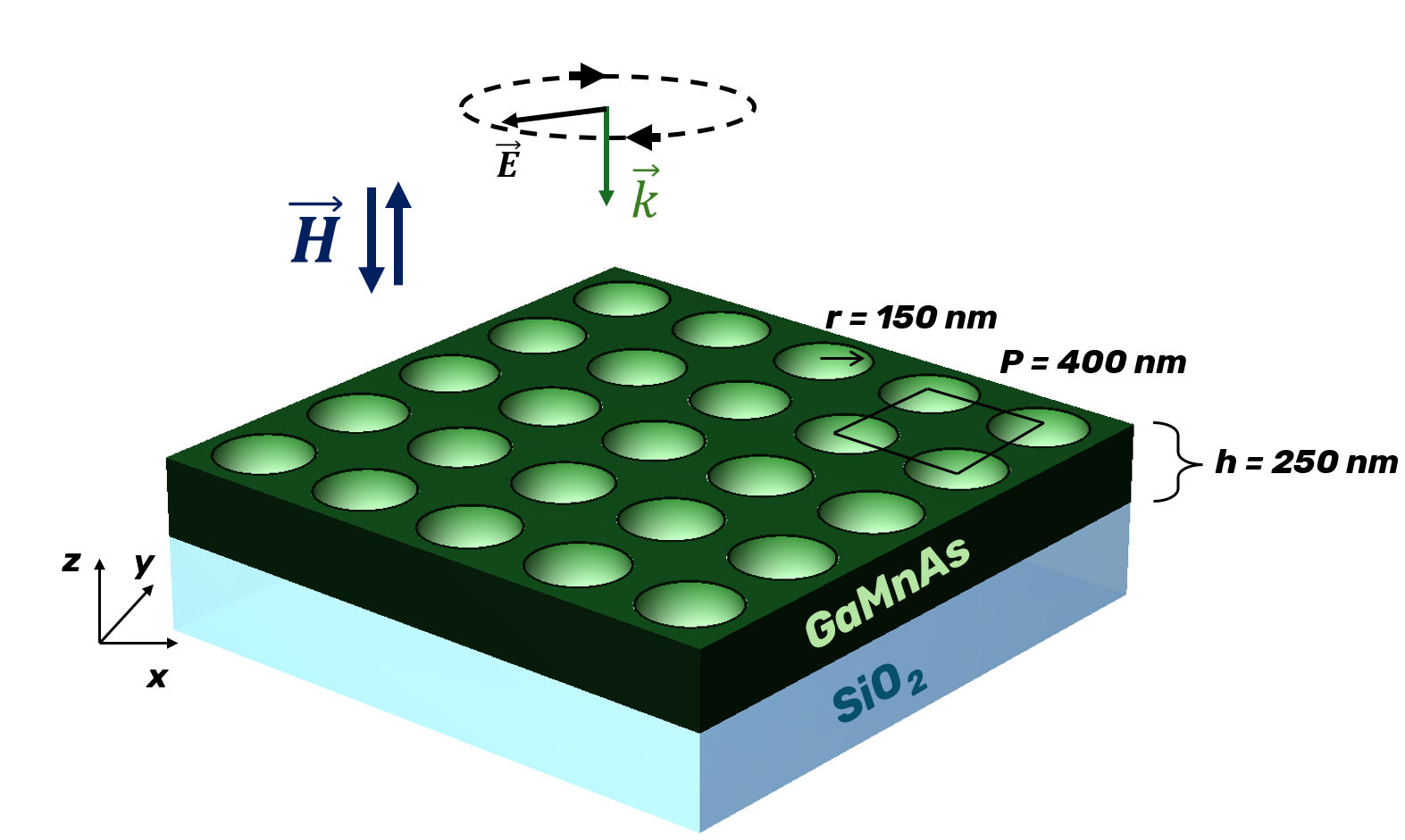}
b)\includegraphics[width=0.4\linewidth]{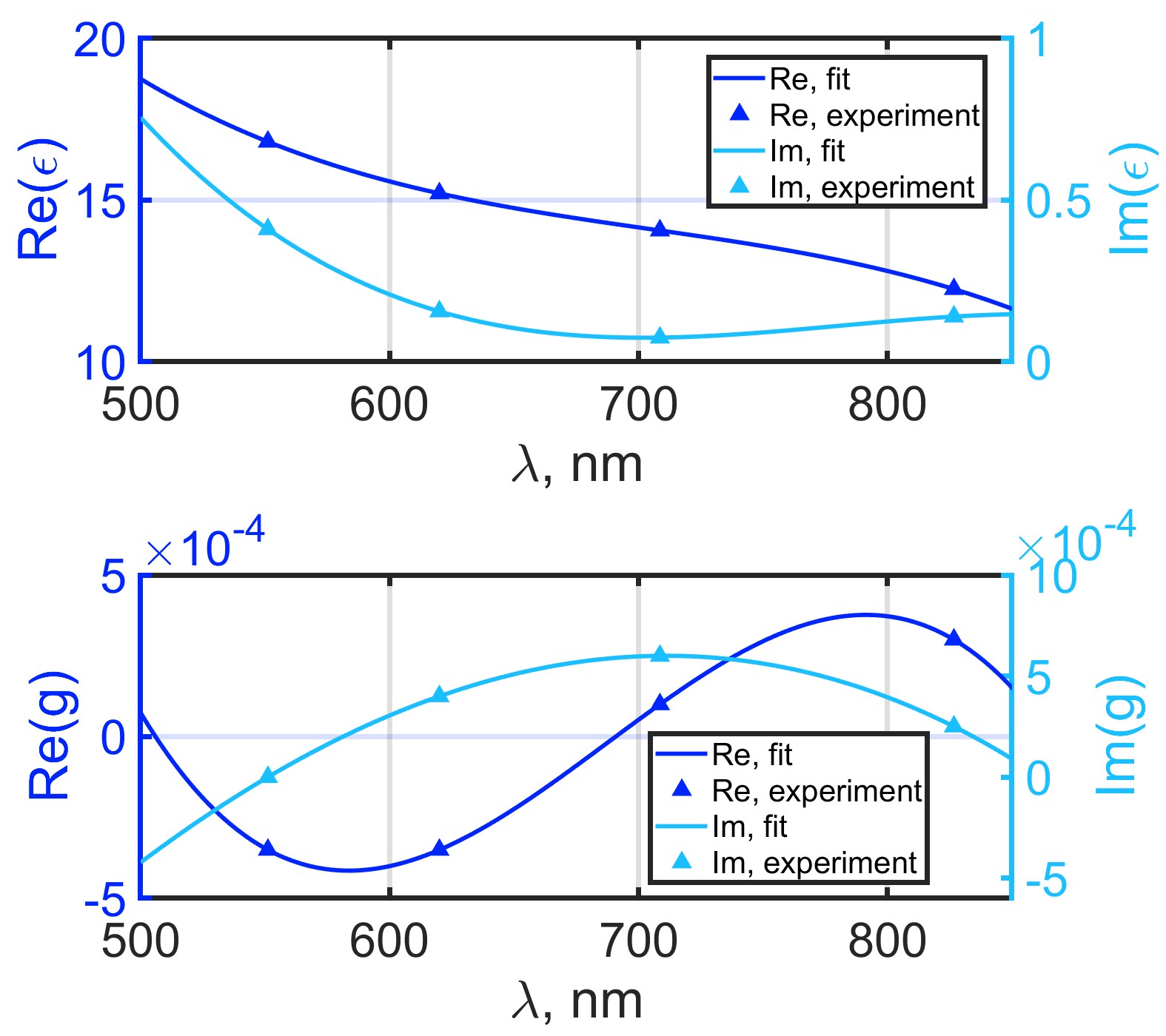}
\caption{Schematic of the considered metasurface (a) and optical and magneto-optical properties of GaMnAs used in the calculations (b).}
\label{fig:geometry}
\end{figure}

In this work, we consider an all-dielectric perforated metasurface based on the magnetic semiconductor GaMnAs (Fig.~\ref{fig:geometry}(a)). The metasurface is formed by a $h=250$~nm thick GaMnAs film perforated by a square lattice of cylindrical air holes extending through the entire layer. The lattice period is $P=400$~nm, and the hole radius is $r=150$~nm. These geometrical parameters were selected by a targeted tuning of the metasurface geometry to spectrally align the resonantly enhanced transmission modulation $\Delta T$ with a high-transmission state and to obtain a pronounced enhancement relative to the smooth GaMnAs film. For the chosen geometry, the enhanced $\Delta T$ occurs in the vicinity of the transmittance maximum, while the multipolar analysis presented below identifies this high-transmission resonant state as an anapole.

The perforated GaMnAs layer is assumed to be placed on a SiO$_2$ substrate. A smooth GaMnAs film of the same thickness on the same SiO$_2$ substrate was also considered as a reference structure. The choice of GaMnAs is motivated by the need to combine the conditions required for anapole excitation with an intrinsic magneto-optical response. Anapole states are typically supported by high-index dielectric nanostructures, whereas the magneto-optical intensity effect requires a nonzero gyration parameter. Iron-garnet films are widely used in magneto-optics; however, their dielectric permittivity is lower than that of some semiconductor materials. GaMnAs therefore provides a suitable platform, combining a relatively high dielectric permittivity with intrinsic magneto-optical activity.

Optical and magneto-optical properties of GaMnAs (Fig.~\ref{fig:geometry}(b)) were described by the dielectric permittivity tensor
\begin{equation}
\hat{\varepsilon}=\begin{pmatrix}
\varepsilon & i g & 0\\
-i g & \varepsilon & 0\\
0 & 0 & \varepsilon
\end{pmatrix},
\label{eq:eps_tensor}
\end{equation}
where $\varepsilon$ is the diagonal dielectric permittivity and $g$ is the gyration parameter. For the corresponding homogeneous medium with tensor~\eqref{eq:eps_tensor}, light propagating along the magnetization direction has two circularly polarized eigenmodes. In the weak-gyrotropy and weak-absorption approximation, $|g|\ll|\varepsilon|$ and $n''\ll n'$, the complex refractive indices of the two circular eigenmodes can be written as \cite{Zvezdin1997ModernMagnetooptics}
\begin{equation}
n_{\pm}=n'\pm\frac{g'}{2n'}+i\left(n''\pm\frac{g''}{2n'}\right),
\label{eq:npm}
\end{equation}
where $n'$ and $n''$ are the real and imaginary parts of the complex refractive index in the absence of gyration, $\varepsilon=(n'+in'')^2$, while $g'$ and $g''$ are the real and imaginary parts of the gyration parameter.

The gyration parameter is an odd function of magnetization: when the magnetization direction $\mathbf{M}$ is reversed, its sign changes, i.e., $g(-\mathbf{M})=-g(\mathbf{M})$. The wave numbers corresponding to the two circular polarizations are defined as $k_{\pm}=k_0n_{\pm}$, where $k_0$ is the wave number in vacuum. Thus, magnetization reversal changes the complex refractive index experienced by a fixed circular polarization and, consequently, the conditions for light transmission through the resonant structure. Equivalently, for a fixed magnetization direction, the two opposite circular polarizations have different complex refractive indices, which gives rise to different transmittances $T_+$ and $T_-$. 

The spectral dependences of the diagonal permittivity $\varepsilon$ and the gyration parameter $g$ were reconstructed from the experimentally determined optical and magneto-optical properties of Ga$_{0.92}$Mn$_{0.08}$As \cite{Terada2015GaMnAs}. These optical and magneto-optical data were obtained at a temperature of 5~K under an external magnetic field of 1~T applied perpendicular to the film plane.

\subsection{Magneto-optical transmission calculations}

To quantify the magneto-optical response in transmission, we consider the change in the transmittance of a fixed incident circular polarization upon magnetization reversal. The transmission modulation is defined as
\begin{equation}
\Delta T=T(+M_z)-T(-M_z),
\label{eq:deltaT}
\end{equation}
where $T(+M_z)$ and $T(-M_z)$ are the transmittances for the magnetization directed along the positive and negative $z$ directions, respectively, with the $z$ axis normal to the metasurface plane.

For comparison with the absolute transmission modulation, we also consider the normalized magneto-optical intensity effect
\begin{equation}
\delta=\frac{\Delta T}{T_{\mathrm{avg}}},
\label{eq:delta}
\end{equation}
where
\begin{equation}
T_{\mathrm{avg}}=\frac{T(+M_z)+T(-M_z)}{2}.
\label{eq:Tavg}
\end{equation}

At normal incidence, these magnetization directions are parallel and antiparallel to the incident wave vector. The transmittances were calculated using rigorous coupled-wave analysis (RCWA). Related scattering-matrix Fourier-modal formulations are well established for calculating the transmission and resonant optical response of finite-thickness two-dimensionally periodic photonic slabs \cite{Tikhodeev2002Quasiguided}. Magnetization reversal was implemented by changing the sign of the gyration parameter, $g(\lambda)\rightarrow-g(\lambda)$, in the off-diagonal components of the permittivity tensor, while the diagonal permittivity remained unchanged. For the geometrically achiral structure considered here at normal incidence, this description is equivalent to comparing the transmittances of the two opposite circular polarizations at a fixed magnetization direction.

The considered metasurface is geometrically achiral: neither the circular shape of the holes nor the square lattice introduces an intrinsic chiral response. At normal incidence, its fourfold rotational symmetry is preserved in the Faraday geometry, because the magnetization is directed along the symmetry axis. Consequently, the two circular polarizations remain independent transmission channels \cite{Kaschke2014CircularPolarizers,FernandezCorbaton2013Helicity}, and their transmittance difference originates from the magneto-optical response of GaMnAs and its resonant modification by the perforated geometry. Therefore, $\Delta T$ is governed in this system by the magneto-optical properties of GaMnAs and by the resonant redistribution of the electromagnetic field in the perforated film.

\subsection{Electromagnetic-field and multipole calculations}

The electromagnetic-field distributions and the multipolar contributions to scattering were calculated using the finite-difference time-domain method implemented in Ansys Lumerical. The multipole decomposition was performed for the induced-current distribution within the GaMnAs part of one $P\times P$ metasurface unit cell using the exact current-based expressions for the electric and magnetic multipole moments beyond the long-wavelength approximation \cite{Alaee2018Multipole}. These expressions contain spherical Bessel functions and include higher-order corrections beyond the conventional long-wavelength moments. The chosen unit cell consists of GaMnAs with quarter-circular air-hole sections at its four corners, which together correspond to one complete air hole per unit cell due to the periodicity of the structure.

\section{Results and discussion}

\subsection{Magnetization-induced transmission modulation}

\begin{figure}[htbp]
\centering
a)\includegraphics[width=0.47\linewidth]{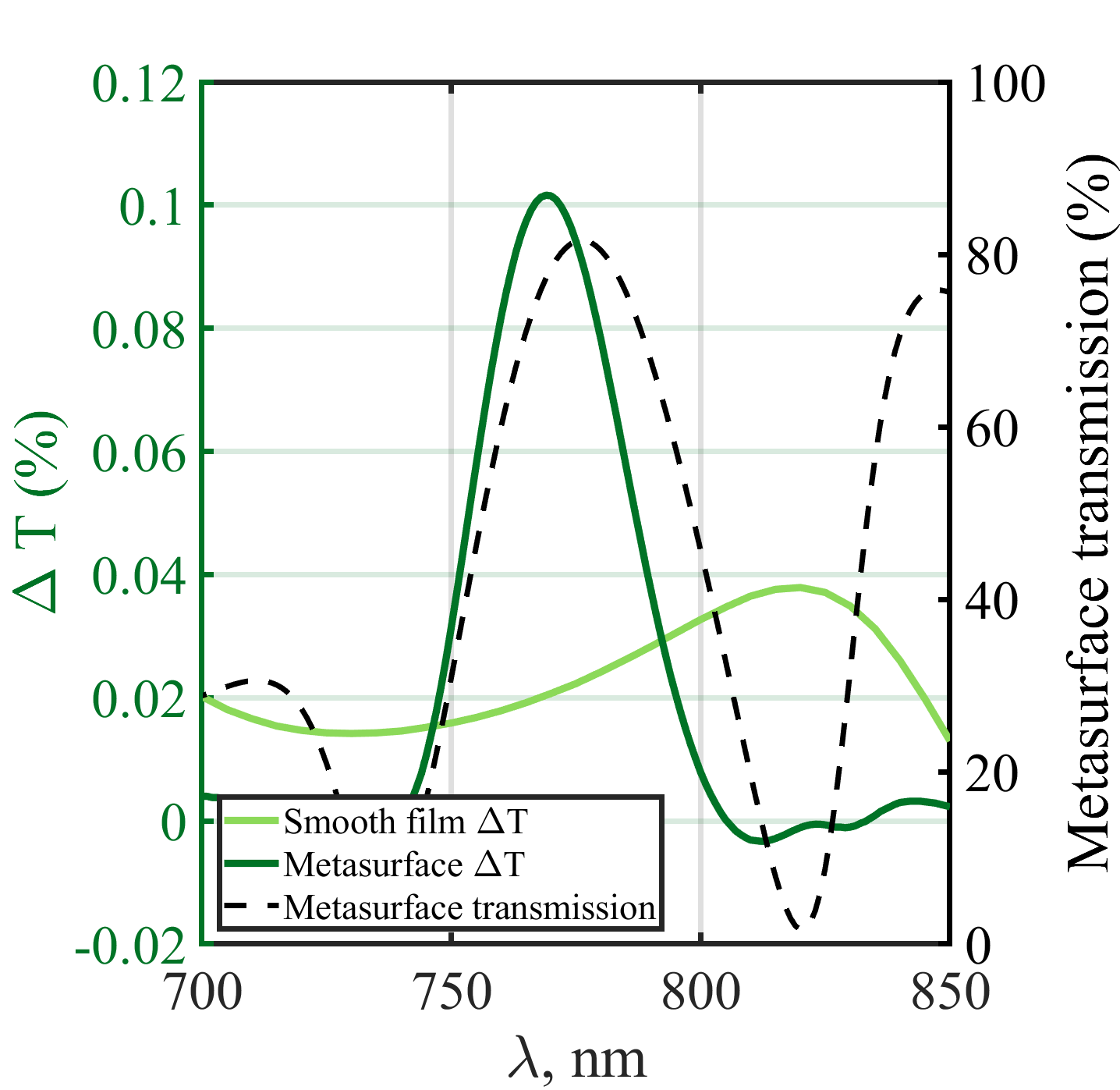}
b)\includegraphics[width=0.47\linewidth]{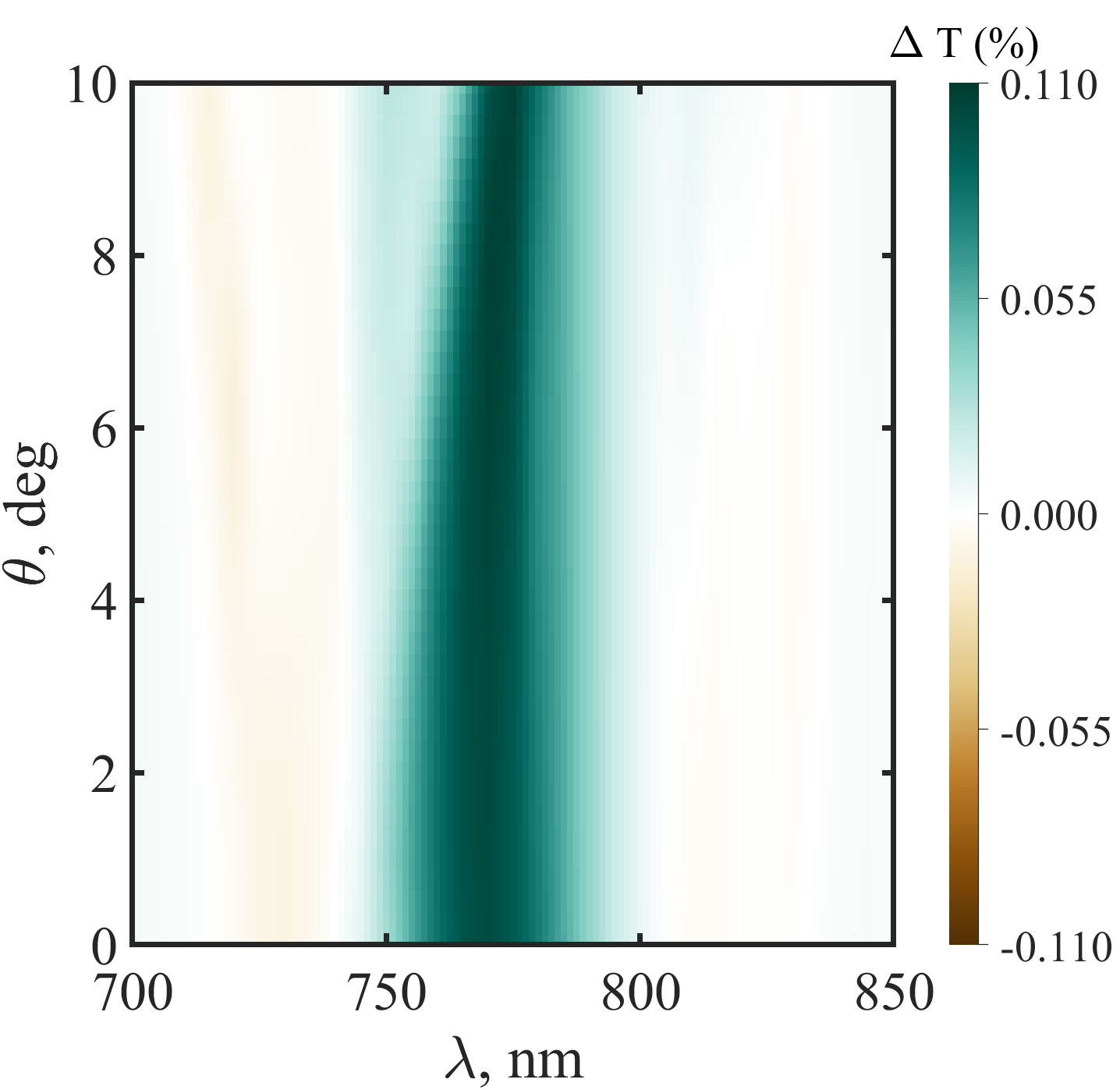}
\caption{Transmission modulation $\Delta T$ of the smooth GaMnAs film and the perforated metasurface. (a) Spectral dependences of the transmission modulation for the smooth film and the perforated metasurface at normal incidence, together with the metasurface transmittance $T$. (b) Angular-spectral map of the transmission modulation $\Delta T$ for the perforated metasurface. The color scale is centered at $\Delta T=0$.}
\label{fig:transmission}
\end{figure}

Figure~\ref{fig:transmission}(a) compares the transmission modulation, $\Delta T$, for the smooth GaMnAs film and the perforated metasurface. In the film $\Delta T$ varies smoothly and remains approximately 0.02\% near $\lambda=770$~nm. In contrast, the perforated structure exhibits a pronounced modulation peak in the same spectral region, where the absolute transmission modulation $\Delta T$ reaches 0.10\%. This value is five times larger than the corresponding transmission modulation of the smooth film. Importantly, this enhancement occurs close to the transmittance maximum of the metasurface, where the transmittance reaches 81\%. Thus, perforation enhances the absolute transmission modulation while maintaining a high-transmission resonant regime. For the angular-dependent calculations shown in Fig.~\ref{fig:transmission}(b), the incidence angle was varied in the $xz$ plane. The enhanced $\Delta T$ region observed near the normal-incidence resonance persists over a broad range of angles and gradually shifts in wavelength as the angle increases. Therefore, the enhanced magneto-optical response is not restricted to strictly normal incidence and remains pronounced under moderately oblique illumination.

\begin{figure}[htbp]
\centering
a)\includegraphics[width=0.47\linewidth]{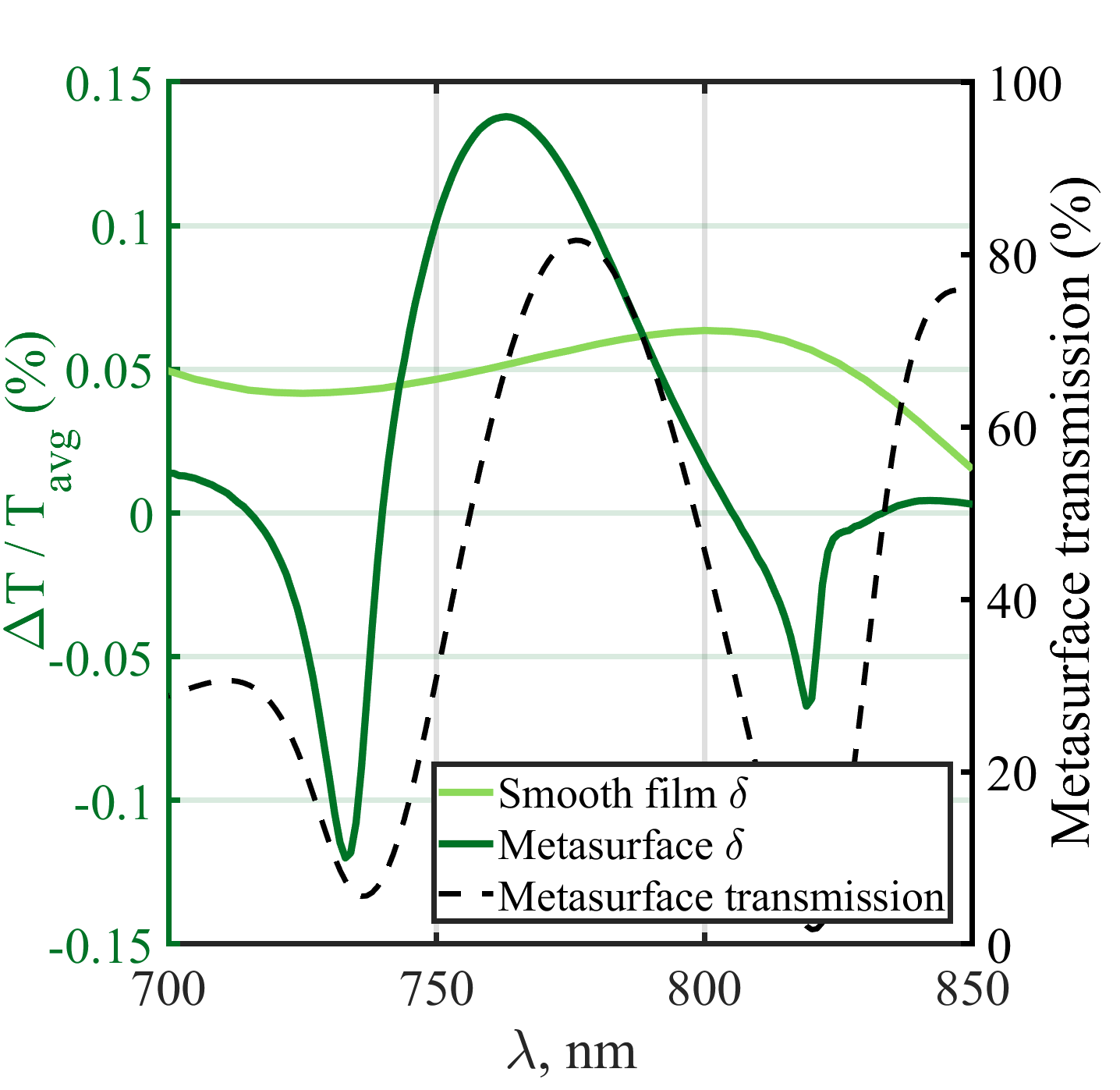}
b)\includegraphics[width=0.47\linewidth]{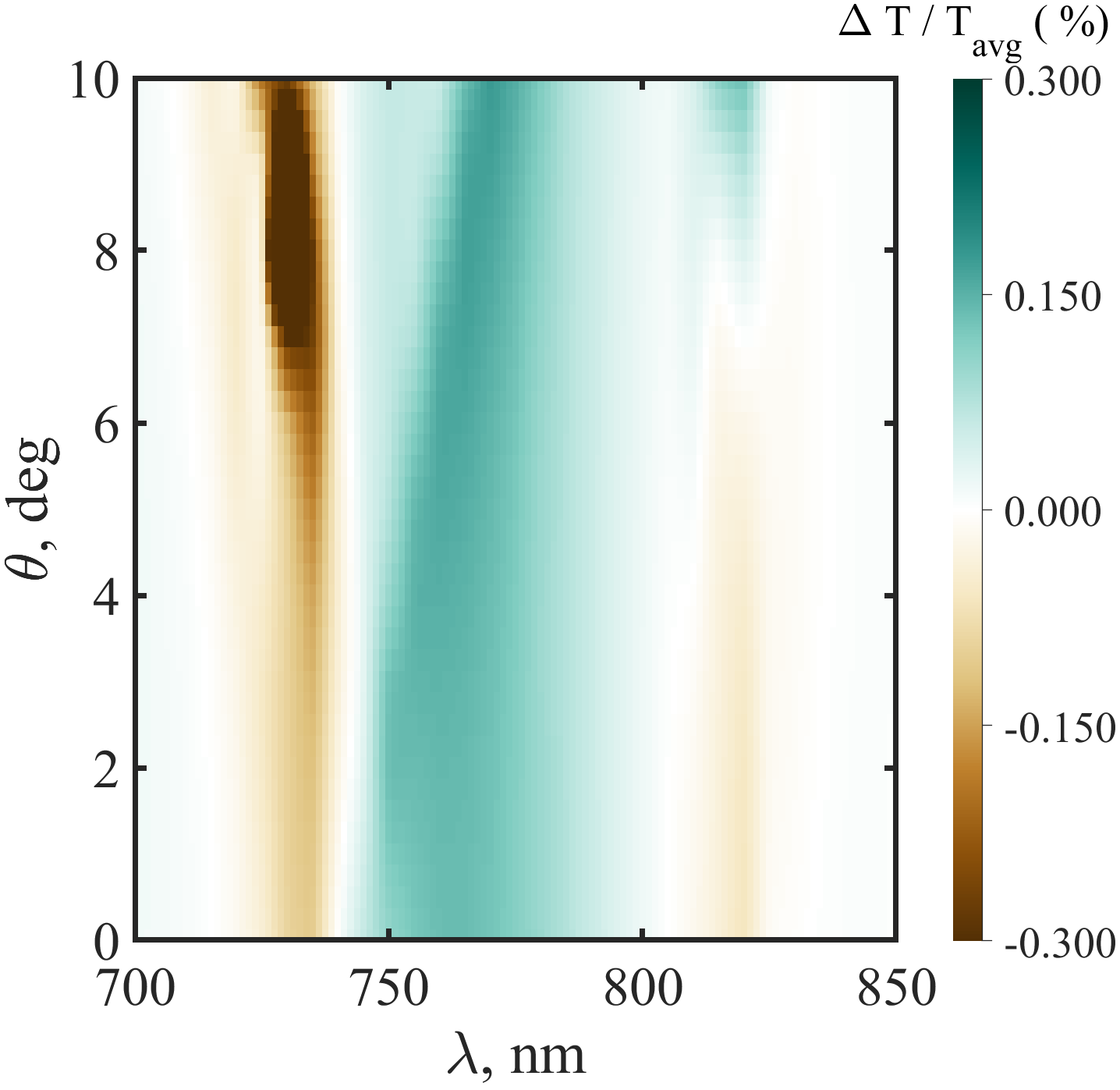}
\caption{Normalized magneto-optical intensity effect $\delta=\Delta T/T_{\mathrm{avg}}$ of the smooth GaMnAs film and the perforated metasurface. (a) Spectral dependences of $\delta$ for the smooth film and the perforated metasurface at normal incidence, together with the metasurface transmittance $T$. (b) Angular-spectral map of $\delta$ for the perforated metasurface. }
\label{fig:normalized_transmission}
\end{figure}

Figure~\ref{fig:normalized_transmission} shows the normalized magneto-optical intensity effect, $\delta=\Delta T/T_{\mathrm{avg}}$. At normal incidence, the maximum value of $\delta$ for the perforated metasurface exceeds that of the smooth GaMnAs film by a factor of 2.72. Compared with the absolute transmission modulation $\Delta T$, the maximum of the normalized effect is spectrally shifted because the normalization involves division by the relatively large transmittance $T_{\mathrm{avg}}$ in the anapole region. As shown in Fig.~\ref{fig:normalized_transmission}(b), the enhanced normalized magneto-optical response persists over the considered incidence-angle range up to $10^\circ$, similarly to the behavior observed for the absolute transmission modulation $\Delta T$.

\subsection{Anapole state identification and enhancement mechanism}

\begin{figure}[htbp]
\centering
a)\includegraphics[width=0.47\linewidth]{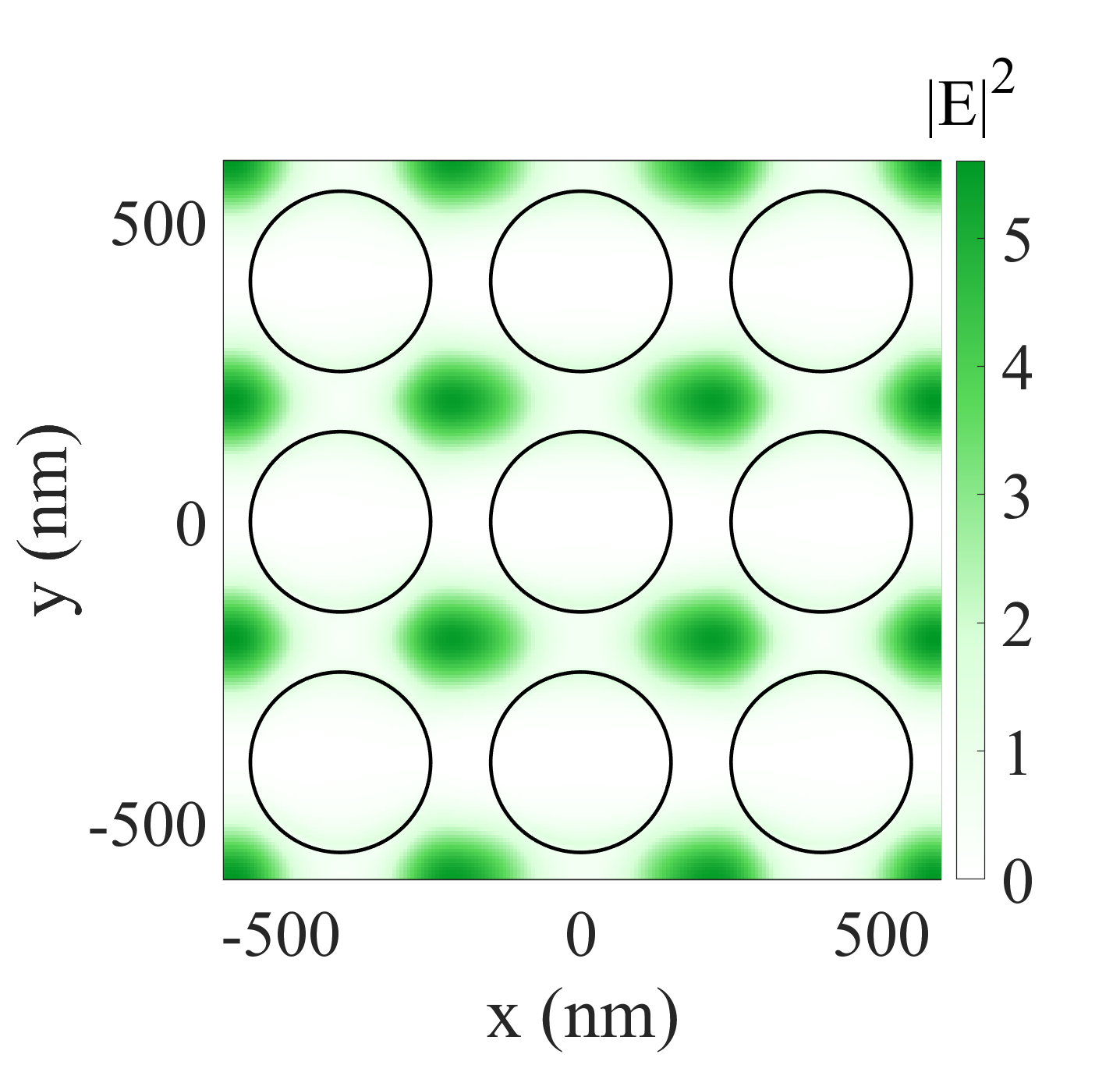}
b)\includegraphics[width=0.47\linewidth]{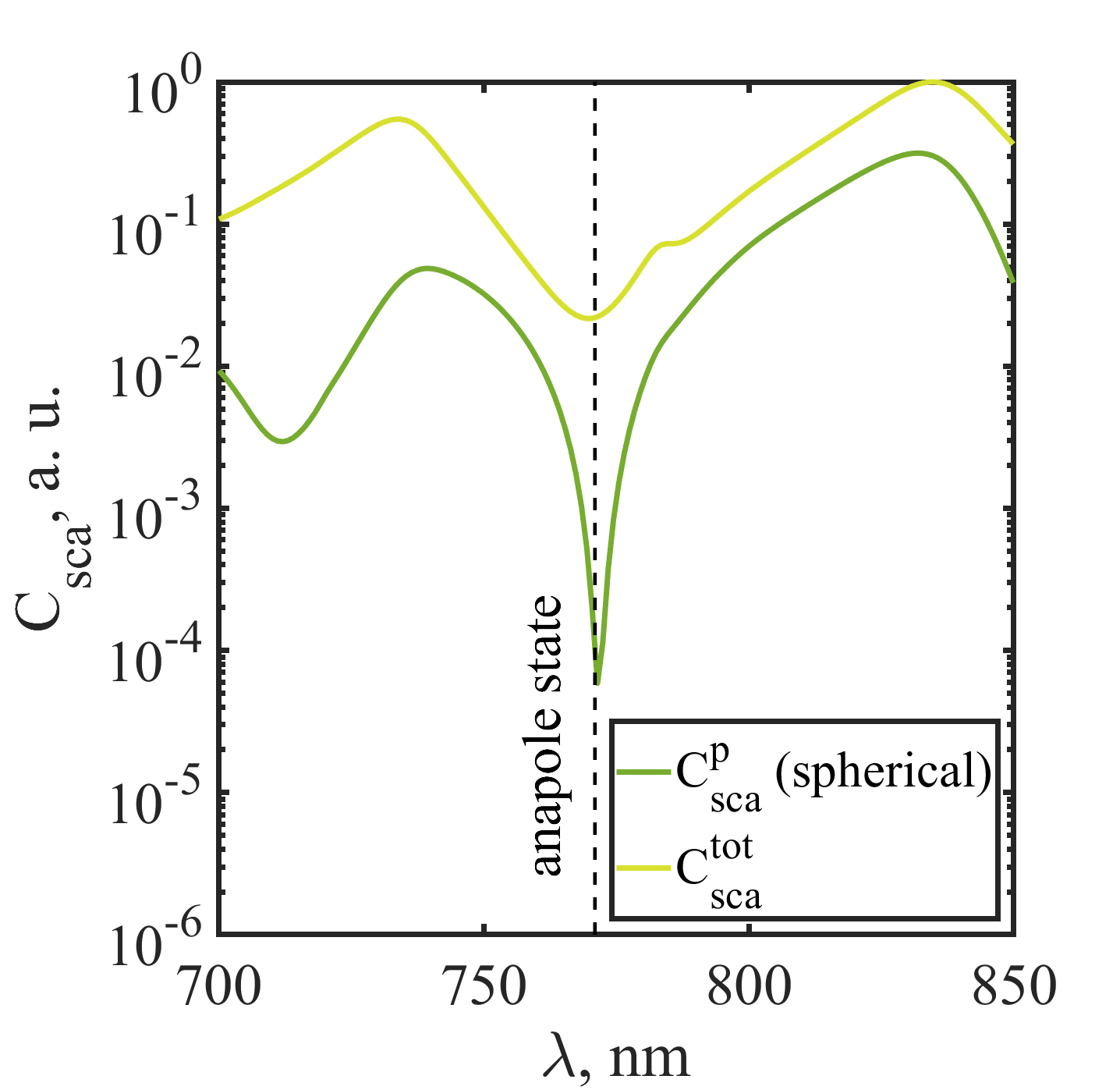}
\caption{(a) Distribution of the electric-field intensity $|E|^2$ in the plane $z=h/2$, corresponding to the mid-height of the perforated GaMnAs layer at $\lambda=770$~nm; black circles indicate the air holes. (b) Multipolar decomposition of the scattering in the vicinity of the anapole state: summed multipolar scattering contribution $C_{\mathrm{sca}}^{\mathrm{tot}}$ and exact spherical electric-dipole contribution $C_{\mathrm{sca}}^{p}$. The vertical dashed line marks $\lambda=770$~nm, corresponding to the anapole-state wavelength.}
\label{fig:anapole}
\end{figure}

The physical origin of this behavior is clarified by the electromagnetic-field distribution and the multipolar analysis of scattering. As shown in Fig.~\ref{fig:anapole}(a), at $\lambda=770$~nm the electric field is localized predominantly in the GaMnAs regions between the air holes. Thus, the resonant field is concentrated directly in the magneto-optically active material. The spatial localization of the electric field within the GaMnAs layer is further quantified by the electric-field confinement factor $\eta_E$ in Appendix~\ref{app:field_confinement}.

When interpreting the anapole state, it is important to distinguish between the conventional Cartesian current-multipole expansion and the exact multipole formulation used here. In the Cartesian representation, the electric dipole moment $\mathbf{p}$ and toroidal dipole moment $\mathbf{T}$ are treated as separate contributions, and the anapole condition is commonly described by the destructive-interference relation $\mathbf{p}+ik\mathbf{T}\approx0$ \cite{Miroshnichenko2015Anapole}. In the exact formulation, these contributions are contained within the same electric-dipole radiation channel through the spherical Bessel functions entering the exact multipole expressions. Expanding these functions in powers of the dimensionless parameter $kr$ shows that the leading term corresponds to the conventional Cartesian electric dipole, the next nonvanishing term to the toroidal dipole, and the subsequent terms to higher-order corrections \cite{Alaee2018Multipole,Li2018AnapoleCondition}. Therefore, suppression of the complete electric-dipole contribution accounts not only for the lowest-order electric-dipole and toroidal-dipole terms but also for the higher-order corrections. Such an identification of anapole states through the suppression of the corresponding electric-dipole contribution to scattering has also been employed in previous studies \cite{Gurvitz2019Toroidal,Tian2019ActiveAnapole}.

As shown in Fig.~\ref{fig:anapole}(b), at $\lambda=770$~nm the electric-dipole contribution $C_{\mathrm{sca}}^{p}$ exhibits a narrow and deep minimum. In the same spectral region, the summed multipolar scattering contribution $C_{\mathrm{sca}}^{\mathrm{tot}}$ shows a local minimum. The remaining scattering contribution arises from other multipolar channels that are not simultaneously suppressed at this wavelength. The pronounced suppression of the electric-dipole contribution, accompanied by a local minimum of the summed multipolar scattering contribution and the strong field localization shown in Fig.~\ref{fig:anapole}(a), identifies this spectral feature as an anapole state. In the present structure, the resonant field associated with the anapole state is localized directly in the magneto-optically active GaMnAs. Thus, the resonant field localization and the magneto-optical response occur within the same material volume. The resulting strong spatial overlap between the resonant field and the gyrotropic medium provides a more favorable condition for an enhanced magneto-optical intensity effect, consistent with the enhanced $\Delta T$ shown in Fig.~\ref{fig:transmission}.

At the same time, according to Eq.~\eqref{eq:npm}, the wave numbers for the two circular polarizations, $k_{\pm}(\mathbf{M})=k_0n_{\pm}(\mathbf{M})$, depend on the magnetization orientation through the sign of the gyration parameter. Therefore, magnetization reversal changes the excitation conditions of the resonant state and the transmission through the structure. In the anapole-state region, where high transmittance is preserved and the field is localized in GaMnAs, this leads to an enhanced transmission modulation $\Delta T$, as defined in Eq.~\eqref{eq:deltaT}.

\section{Conclusions}

We have shown that perforating a GaMnAs film enhances the magneto-optical intensity effect while preserving high optical transparency. Near the anapole state, the absolute transmission modulation for a fixed incident circular polarization increases by a factor of five compared with the smooth film, while the metasurface transmittance reaches 81\%.

The enhancement originates from electromagnetic field localization in the GaMnAs regions between the air holes and from suppression of the radiating electric-dipole channel in the anapole regime. Thus, the magneto-optical response is enhanced not at the expense of the transmitted signal, but in a transparent anapole regime.

An important advantage of the proposed system is its simple all-dielectric geometry: a single perforated magnetic film simultaneously supports the resonant state and has a nonzero gyration parameter, without the need to combine several materials in a hybrid structure. The use of experimentally determined GaMnAs parameters makes the considered metasurface close to a realistic structure and promising for future experimental verification.

The obtained results may be useful for the development of compact magneto-optical elements for controlling circularly polarized light in transmission.

\appendix

\section{Electric field confinement in the GaMnAs layer}
\label{app:field_confinement}

To quantify the spatial localization of the electric field within the magneto-optically active material, we introduce the electric field confinement factor
\begin{equation}
\eta_E(\lambda)=
\frac{
\displaystyle\int_{V_{\mathrm{GaMnAs}}}|E(\mathbf{r},\lambda)|^2\,dV
}{
\displaystyle\int_{V_{\mathrm{cell}}}|E(\mathbf{r},\lambda)|^2\,dV
},
\label{eq:etaE}
\end{equation}
where $V_{\mathrm{GaMnAs}}$ denotes the GaMnAs part of the patterned layer and $V_{\mathrm{cell}}$ is the entire $P\times P\times h$ volume of one metasurface unit cell, including both the GaMnAs and air-hole regions. The quantity $\eta_E$ therefore characterizes the fraction of the integrated electric-field intensity located within the magneto-optically active GaMnAs material.

\begin{figure}[htbp]
\centering
\includegraphics[width=0.85\linewidth]{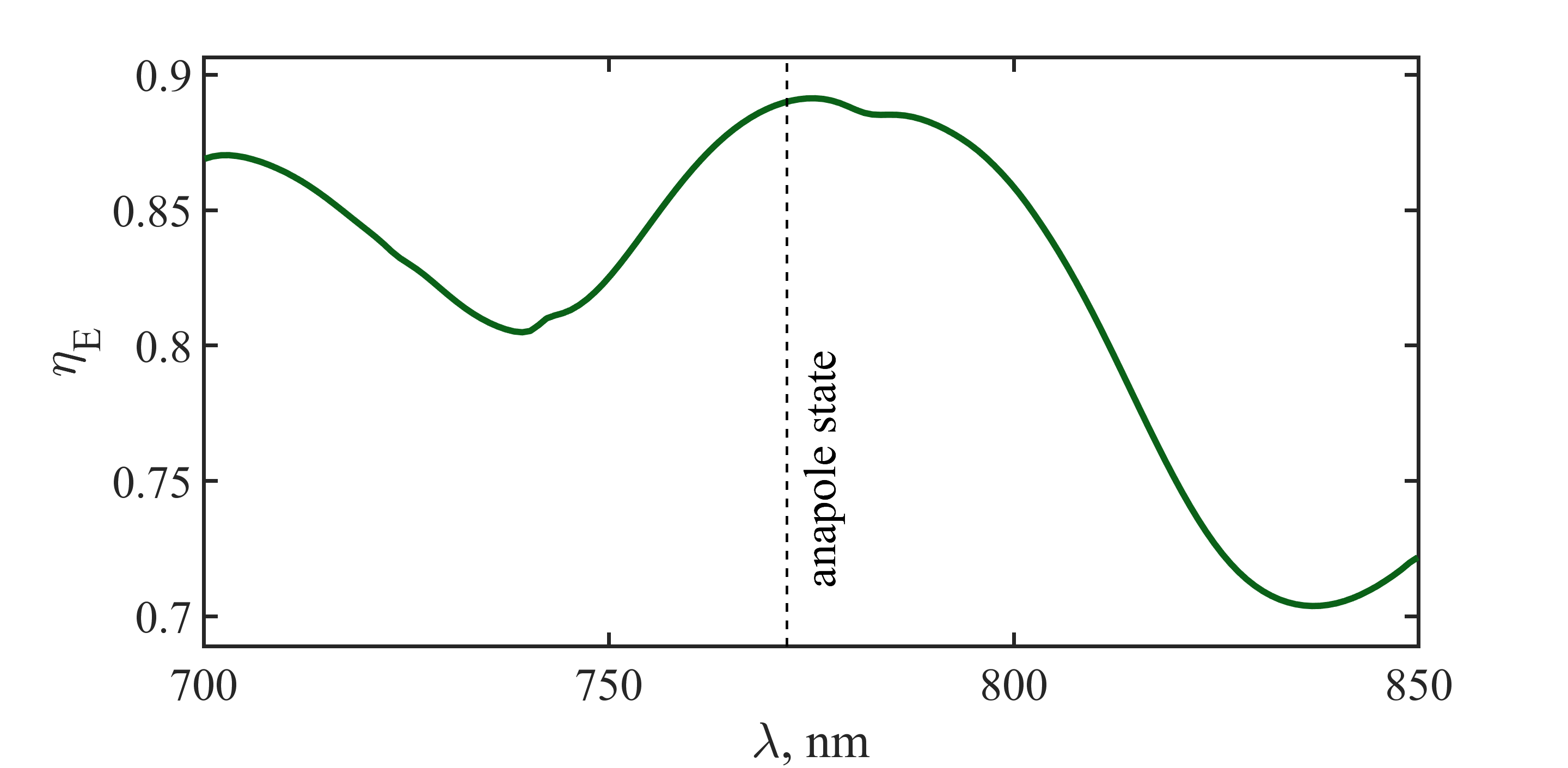}
\caption{Spectral dependence of the electric-field confinement factor $\eta_E$ in the perforated GaMnAs metasurface. The dashed vertical line marks the anapole state wavelength at $\lambda=770$~nm.}
\label{fig:etaE}
\end{figure}

Figure~\ref{fig:etaE} shows the spectral dependence of $\eta_E$ in the wavelength range of interest. The confinement factor increases toward the anapole spectral region and reaches a local maximum close to $\lambda=770$~nm, where approximately 89\% of the integrated electric field intensity within the patterned unit cell is located in GaMnAs. The increase of $\eta_E$ toward the anapole region occurs in the same resonant spectral range in which the normalized magneto-optical intensity effect is enhanced (Fig.~\ref{fig:normalized_transmission}(a)).

\begin{backmatter}

\bmsection{Author contributions}
Polina V. Zorina: Methodology, Software, Investigation, Formal analysis, Visualization, Writing--original draft. Andrey N. Kalish: Supervision, Validation, Writing--review and editing. Vladimir I. Belotelov: Supervision, Writing--review and editing. Venu Gopal Achanta: Supervision. Daria O. Ignatyeva: Conceptualization, Methodology, Supervision, Project administration, Writing--review and editing.

\bmsection{Funding}
This work was supported by the Russian Science Foundation 24-42-02008.


\bmsection{Disclosures}
The authors declare no conflicts of interest.

\bmsection{Data availability}
The data supporting the findings of this study are available from the corresponding author upon reasonable request.

\end{backmatter}


\bibliography{ga_mnas_josab_references}

\end{document}